\documentstyle[amssymb,manuscript,aps,epsf]{revtex}

\begin{document}
\draft
\author{Serge Winitzki\thanks{%
email: S.Winitzki@damtp.cam.ac.uk}}
\address{DAMTP, University of Cambridge, Cambridge CB3 9EW, UK}
\title{Non-Gaussian signature of weak gravitational lensing in the CMB}
\date{June 7, 1998}
\maketitle

\begin{abstract}

Inflationary cosmologies predict Gaussian primordial fluctuations, but
subsequent gravitational lensing of the CMB disturbs its Gaussianity.
Knowledge of the specific signature of lensing is necessary to distinguish
a lensed Gaussian sky from an intrinsically non-Gaussian one. In this paper
we investigate the pattern of non-Gaussian features in the CMB resulting
from gravitational lensing, both analytically and numerically. We describe
the lensed CMB temperature distribution using the formalism of generating
functionals and obtain the complete picture of emerging non-Gaussian
effects. To test for the signature of lensing in practice, we propose a
scale-dependent statistic based on cumulants of the CMB temperature, and we
compute that statistic numerically for typical inflationary models. The effect
is expected to be most significant at angular scales $\sim 10^\prime$
and its magnitude is within observational capabilities of future
satellite missions.

\end{abstract}


\section{Introduction}

The cosmic microwave background (CMB) radiation is a direct imprint of
the primordial matter, and its importance for cosmology cannot be
overestimated \cite{CMB}. The forthcoming MAP \cite{MAP} and PLANCK
\cite{PLANCK} satellite missions are expected to carry out high-resolution,
low-noise measurements of the CMB temperature and polarization, yielding a
stringent test of our theories of cosmological structure formation. One of
the pronounced differences between the competing theories is a Gaussian
distribution of the density fluctuations in most inflationary theories
versus the generically non-Gaussian inhomogeneities in
scenarios based on topological defects. For this reason, detection of
non-Gaussian features in the CMB is important for distinguishing the
cosmological models.

However, even if the fluctuations of the CMB were Gaussian at the surface
of last scattering, their later evolution would have introduced certain
non-Gaussian features in their distribution. One of such effects is
gravitational lensing by matter present along trajectories of the
primordial photons. While the effect of weak gravitational lensing on the
power spectrum of the CMB fluctuations has been extensively studied (see
e.g.\ \cite{Seljak96} and references therein), including its relevance to
observations of large-scale structure \cite{Spergel} and cosmological
parameter estimation \cite{Efstathiou98}, it is only recently that
attention has been drawn to the appearance of non-Gaussian signal in the
CMB due to weak lensing \cite {Bernardeau96,Bernardeau}.

To identify a sky map as coming from an intrinsically non-Gaussian model
rather than from a lensed inflationary one, we need to distinguish the
non-Gaussian signature of lensing from that of other processes. The goal of
this paper is to investigate in detail the effect of weak gravitational
lensing on the distribution of the CMB fluctuations and to fully
characterize the emergence of non-Gaussian features. Rather than restrict
ourselves to a specific criterion of Gaussianity at the outset, we
describe the lensed sky in a general way by using the formalism of
generating functionals, somewhat similar to that used in quantum field
theory. A description of a random field in terms of a generating functional
is equivalent to giving the full probability distribution and in principle
is enough to evaluate any field statistic.

To test for the signature of
lensing in practice, we chose a criterion based on cumulants of the temperature
distribution \cite{Cumulants}. Cumulants are generic additive and
scale-sensitive probes of Gaussianity defined through moments of the field
and are easy to compute in our context. Recent studies of geometrical
criteria of Gaussianity \cite{MFPaper} show that the
``generalized genus'' statistics based on Minkowski functionals \cite
{MFOthers} are sensitive to the $3$-point temperature distribution
at points separated by the smallest map scale. A direct examination of that
temperature distribution using cumulants yields a statistic essentially
equivalent to or possibly more discriminative than the generalized genus.
Thus motivated, we apply a cumulant statistic to detect the
non-Gaussian effects of lensing.

The distribution of the CMB after lensing depends on the initial CMB
power spectrum and on the distribution and evolution of the lensing
sources, all of which have to be taken from a particular cosmological
model. We compute the effects of lensing in typical inflationary CDM
cosmological models, using the program {\tt CMBFAST} \cite{CMBFAST} to
obtain the unlensed CMB power spectrum and the lens distribution. As our
Gaussianity test, we evaluate the dimensionless cumulant $\bar{\chi}_{22}(
\theta )$ which of all cumulants is expected to show the largest
deviation from the Gaussian value $\bar{\chi}_{ij}=0$, as well as
the next largest cumulant $\bar{\chi}_{33}$.
Using a wide range of cosmological parameters, we find that
the maximum signal as measured by $\bar{\chi}_{22}\left( \theta \right) $
is generically expected to be of order $10^{-2}$-$10^{-3}$ at angular
scales $\theta \sim 10$-$20$ arcmin. The value of $\bar{\chi}_{33}$ is
smaller by several orders of magnitude.

The paper is laid out as follows: In the next section we present an outline
and the results of our main calculation of the generating functional for the
temperature distribution after lensing, in a model with Gaussian initial
fluctuations and Gaussian distribution of lensing sources. The detailed
definitions of the formalism are found in Appendix \ref{App:WeaklynonG},
while the calculation itself is deferred to Appendix \ref
{App:GenFunc}. We also comment on a generalization of our formalism to
non-Gaussian initial conditions and lens distributions.
In Section \ref{Sec:Observable} and Appendix \ref{App:Chi} we
compute the observable effect of lensing as measured by the cumulant $\bar{%
\chi}_{22}\left( \theta \right) $ and investigate the possibility to detect
this signal, which may just be at the lower limit of angular resolution and
sensitivity of the forthcoming PLANCK mission.

\section{Non-Gaussian effects of lensing on the CMB}

\label{Sec:NonGL}

To achieve a clear understanding of non-Gaussianity due to lensing, we first
wish to concentrate on the simplest case when both the initial CMB
fluctuations and the lensing sources are Gaussian. The effect of lensing is
to make a primordial photon appear to originate from a slightly different
direction of the sky. Accordingly, one
uses the ``lensing field'' ${\bf g}\left( {\bf x}\right) $ to shift the
original temperature field $T_0\left( {\bf x}\right) $, so that after
lensing it becomes the observed $T\left( {\bf x}\right) $,
\begin{equation}
T\left( {\bf x}\right) \equiv T_0\left( {\bf x}+{\bf g}\left( {\bf x}\right)
\right) .  \label{LensedField}
\end{equation}
The temperature before lensing is taken as a homogeneous Gaussian random
field $T_0\left( {\bf x}\right) $ with zero mean and known correlation
function
\begin{equation}
\left\langle T_0\left( {\bf x}\right) T_0\left( {\bf y}\right) \right\rangle
\equiv C_0\left( {\bf x}-{\bf y}\right) .
\end{equation}
(Here ${\bf x}$ is a point in a region of the $2$-dimensional Euclidean
space ${\bf R}^2$ which represents a small flat patch of the sky. As we
shall see below, all lensing effects appear on sub-degree scales, and the
small-angle approximation used throughout this paper suffices for our
purposes.) The ``lensing field'' ${\bf g}\left( {\bf x}\right) $ is also
assumed to be a homogeneous zero-mean Gaussian random function on ${\bf R}^2$
with values in the same space ${\bf R}^2$ and known (matrix-valued)
correlation function
\begin{equation}
{\bf C}_g\left( {\bf x}-{\bf y}\right) \equiv {\bf C}_g^{kl}\left( {\bf x}-%
{\bf y}\right) =\frac 12\delta ^{kl}C_g\left( {\bf x}-{\bf y}\right) .
\label{GammaCF}
\end{equation}
(The function $C_g$ is the same as $C_{gl}$ in the notation of Ref.\ \cite
{Seljak96}.) Due to homogeneity, all our correlation functions depend only
on distance, $C\left( {\bf x}\right) =C\left( \left| {\bf x}\right| \right)$.

We shall assume below that the angular displacement ${\bf g}$ due to lensing
is ``small'' and it makes sense to expand in ${\bf g}$. Quantitatively, the
lensing effect is small if the standard deviation of the deflection angle $%
{\bf g}$ gives a scale at which the original temperature field does not
change appreciably. This condition can be expressed through the real-space
correlation functions of the two fields as
\begin{equation}
C_g\left( 0\right) \ll \frac{C_0\left( 0\right) }{\left| C_0^{\prime \prime
}\left( 0\right) \right| }.  \label{CGSmall}
\end{equation}
As we verified numerically, this condition holds for all inflationary
models. (The values of $C_g\left( 0\right) $ are $\lesssim 10^{-7}$ while
the right hand side of Eq.\ (\ref{CGSmall}) is typically $\sim 10^{-5}$.)

The correlation functions $C_0\left( x\right) $ and $C_g\left( x\right) $
can be computed from any given cosmological model. The CMB correlation
function is conventionally expressed through the angular power spectrum $C_l$%
,
\begin{equation}
C_0\left( \theta \right) =\frac 1{4\pi }\sum_{l=0}^\infty \left( 2l+1\right)
C_lP_l\left( \cos \theta \right) .  \label{CxFromCl}
\end{equation}
The angular displacement ${\bf g}\left( {\bf x}\right) $ is found by
integrating the transverse gradient of the gravitational potential $\phi $
from the time of last scattering to the present time. The gravitational
potential is related to the fluctuation of the primordial matter density $%
\delta \rho $ by a Poisson equation. The explicit expressions are derived by
several authors (see e.g.\ \cite{Seljak94,Seljak96}) and we shall not need
to reproduce them here. Although the distribution of density fluctuations $%
\delta \rho $ and hence of the gravitational potential $\phi $ may be
non-Gaussian due to either primordial topological defects or nonlinear
evolution, the averaging over photon trajectories is expected to wash out,
due to the central limit theorem, at least some of the non-Gaussianity. We
shall comment on more general initial conditions later, and for now we shall
assume that a certain model of cosmological structure formation has been
chosen and the corresponding correlation matrix ${\bf C}_g$ of ${\bf g}%
\left( {\bf x}\right) $ has been computed.

Even if both the original field $T_0\left( {\bf x}\right) $ and the
deflection ${\bf g}$ are Gaussian, the field $T\left( {\bf x}\right) $ after
lensing has a non-Gaussian distribution. We shall describe the distribution
of $T\left( {\bf x}\right) $ through its generating functional $Z\left[
J\right] $ defined as the average
\begin{equation}
Z\left[ J\right] \equiv \left\langle \exp \left[ -i\int J\left( {\bf x}%
\right) T\left( {\bf x}\right) d{\bf x}\right] \right\rangle _{T\left( {\bf x%
}\right) }\equiv \left\langle \exp \left[ -iJ_xT_x\right] \right\rangle _T,
\label{ZDef1}
\end{equation}
which depends on the parameter function $J({\bf x})$. The
generating functional of a Gaussian distribution must be a Gaussian
functional, such as
\begin{equation}
Z_G\left[ J\right] =\exp \left[ -\frac 12\int J({\bf x})J({\bf y})
C({\bf x},{\bf y}) d{\bf x}d{\bf y}\right]
\equiv \exp \left[ -\frac 12 J_xJ_yC^{xy}\right] .
\end{equation}
A distribution with slight deviations from Gaussianity may be described by a
generating functional of the form
\begin{equation}
Z\left[ J\right] =\exp \left[ -\frac 12J_xJ_yC^{xy}+\frac{\left( -i\right) ^3%
}{3!}J_xJ_yJ_zC_3^{xyz}+...\right] ,  \label{ZDefC1}
\end{equation}
where the Gaussian (quadratic in $J$) term in the exponential still gives the
correlation function $C\left( {\bf x}-{\bf y}\right) $, while the
coefficients $C_n$, $n=3,4,...$ of the non-Gaussian terms are the general
cumulants (see Appendix \ref{App:WeaklynonG} for a more detailed
exposition). Taken together, the cumulants fully characterize the deviations
from a Gaussian distribution, and to find them or equivalently the
functional of Eq.\ (\ref{ZDefC1}) is the main goal of this section.

We now give a sketch of the calculation of $Z\left[ J\right] $ for the
lensed temperature field; the full details are found in Appendix \ref
{App:GenFunc}. Combining Eqs.\ (\ref{LensedField}) and (\ref{ZDef1}) gives
\begin{equation}
Z\left[ J\right] =\left\langle \exp \left[ -i\int J\left( {\bf x}\right)
T_0\left( {\bf x}+{\bf g}\left( {\bf x}\right) \right) d{\bf x}\right]
\right\rangle _{{\bf g},T_0} ;
\end{equation}
here the average must be performed over all realizations of $T_0({\bf x})$
and ${\bf g}({\bf x}) $. The result can be expressed
through the unmodified correlation function $C_0\left( {\bf x}\right) $ and
the generating functional
\begin{equation}
Z_g\left[ {\bf H}\right] \equiv \exp \left[ -\frac 12{\bf H}_x^T{\bf C}%
_g^{xy}{\bf H}_y\right]
\end{equation}
of the deflection field ${\bf g}$, and we obtain (cf.\ Eq.\ (\ref{Ans1}))
\begin{equation}
Z\left[ J\right] =\exp \left[ -\frac 12J_xC_0\left( {\bf x}-{\bf y}+i\frac
\delta {\delta {\bf H}\left( {\bf x}\right) }-i\frac \delta {\delta {\bf H}%
\left( {\bf y}\right) }\right) J_y\right] _{{\bf H}=0}Z_g\left[ {\bf H}%
\right] .  \label{ZExp1}
\end{equation}
We are actually more interested in finding the logarithm of $Z\left[ J\right]
$ since, as Eq.\ (\ref{ZDefC1}) illustrates, $\ln Z\left[ J\right] $ is the
generating function of the cumulants. If one expands $C_0$ in the
exponential of Eq.\ (\ref{ZExp1}),
\begin{eqnarray}
Z\left[ J\right] &=&\exp \left[ -\frac 12J_xC_0^{xy}J_y-\frac{iJ_xJ_y\nabla
C_0\left( {\bf x}-{\bf y}\right) }2\left( \frac \delta {\delta {\bf H}\left(
{\bf x}\right) }-\frac \delta {\delta {\bf H}\left( {\bf y}\right) }\right)
\right.  \nonumber \\
&&\left. +\frac{J_xJ_y\nabla \nabla C_0\left( {\bf x}-{\bf y}\right) }{%
2\cdot 2!}\left( \frac \delta {\delta {\bf H}\left( {\bf x}\right) }-\frac
\delta {\delta {\bf H}\left( {\bf y}\right) }\right) ^2+...\right] _{{\bf H}%
=0}Z_g\left[ {\bf H}\right] ,  \label{ZExp2}
\end{eqnarray}
one obtains an expression formally similar to the generating functional of
Green's functions in interacting quantum field theory with the ``couplings''
represented by the coefficients at the functional derivative terms.
Following the analogy with quantum field theory, we notice that $\ln Z\left[
J\right] $ is the sum of the appropriate connected Feynman diagrams having
no external lines (as shown in Figs.\ 1, 2), with vertices being derivatives
of $C_0\left( {\bf x}\right) $ of all orders and the ``propagator'' being $%
{\bf C}_g\left( {\bf x}-{\bf y}\right) $. The full expression corresponding
to a vertex of arbitrary order is given by Eq.\ (\ref{VertexN}). Unlike the
usual perturbation theory, the order of a diagram is now equal to
the number of propagators in it, because the smallness of the lensing effect
is passed on to the propagator rather than to the vertices. Since the expansion
in Eq.\ (\ref{ZExp2}) is formally to all orders, an infinite number of
vertices with arbitrary number of legs arises, but diagrams with many-legged
vertices are suppressed because they require more propagators and therefore
are of higher order in $C_g$.

Since all vertices in our diagrams are quadratic in $J$, each diagram yields
an even power of $J$. After adding up all diagrams, we therefore expect to
obtain an expression for the generating functional such as
\begin{equation}
\ln Z\left[ J\right] =\frac{i^2}{2!}J_xJ_yC^{xy}+\frac{i^4}{4!}%
J_wJ_xJ_yJ_zC_4^{wxyz}+\frac{i^6}{6!}J_uJ_vJ_wJ_xJ_yJ_zC_6^{uvwxyz}...
\label{LnZAns}
\end{equation}
containing only even powers of $J$, with the even-order cumulants $C$, $C_4$%
, $C_6$, ... found as sums of diagrams with the appropriate power of $J$.
The new correlation function $C\left( {\bf x}\right) $ and the fourth-order
cumulant $C_4$ are (to lowest order in $C_g$) given by the two diagrams in
Fig.\ 1b,
\begin{equation}
C\left( {\bf x}\right) =C_0\left( x\right) +\frac 12\left( C_g\left( 0\right)
-C_g\left( x\right) \right) \left( C_0^{\prime \prime }\left( x\right) +
\frac{C_0^{\prime }\left( x\right) }x\right) +O\left( C_g^2\right) ,
\label{C2Ans}
\end{equation}
\begin{eqnarray}
C_4\left( {\bf w},{\bf x},{\bf y},{\bf z}\right) &=&\frac{1}{4}C_0^{\prime }\left(
\left| {\bf w}-{\bf x}\right| \right) C_0^{\prime }\left( \left| {\bf y}-%
{\bf z}\right| \right) C_g\left( \left| {\bf w}-{\bf y}\right| \right) \frac{%
\left( {\bf w}-{\bf x}\right) \cdot \left( {\bf y}-{\bf z}\right) }{\left|
{\bf w}-{\bf x}\right| \left| {\bf y}-{\bf z}\right| }  \nonumber \\
&&+(\text{23 permutations of }{\bf w}\text{, }{\bf x}\text{, }{\bf y}\text{,
}{\bf z})+O\left( C_g^2\right) .  \label{C4Ans}
\end{eqnarray}
(The expression in Eq.\ (\ref{C4Ans}) is effectively symmetrized in all four
parameters.) The higher-order cumulants are more complicated and we shall
not write them out explicitly.
The functional of Eq.\ (\ref{LnZAns}) together with the diagrammatic method of
computing the coefficients $C$, $C_4$, $C_6$, ... is our main result describing
the characteristic non-Gaussian contributions of gravitational lensing.

From the
spatial dependence of the terms in Eq.\ (\ref{C4Ans}) it is clear that
the $4$-th order cumulant $C_4$ vanishes for certain combinations of points,
for instance $C_4\left( {\bf x},{\bf x},{\bf x},{\bf y}\right) =0$. It can
be shown that similar restrictions hold for higher-order cumulants and, in
particular, that the one-point cumulants $C_n\left( {\bf x},{\bf x},...,
{\bf x}\right) $ for $n>2$ vanish, which means that the one-point temperature
distribution remains Gaussian. This is a manifestation of the specific signature
of gravitational lensing: it only modifies the distribution of relative angular
distances between the CMB photons.

The condition that first-order corrections to the power spectrum due to
lensing are small is, by inspection of Eq.\ (\ref{C2Ans}), equivalent to
Eq.\ (\ref{CGSmall}), so the expansion in $C_g$ is justified.
In typical cosmological models, the
magnitude of $C_g$ is never larger than $10^{-2}$ in
units of the right hand side of Eq.\ (\ref{CGSmall}). Since the total power
of $J$ in a diagram is equal to twice the number of vertices, it follows
that diagrams contributing to higher-order cumulants will be suppressed by
additional factors of $C_g$, and the non-Gaussian signal in higher-order cumulants
rapidly becomes negligible. The most
significant effects of gravitational lensing are therefore
expected in the power spectrum (i.e.\ in $C\left( {\bf x}\right) $) and in
the $4$-th order cumulant $C_4$.

Eqs.\ (\ref{C2Ans})--(\ref{C4Ans}) are consistent with the results of Seljak
\cite{Seljak96} and Bernardeau \cite{Bernardeau}. Our approach is somewhat
similar to that of Ref.\ \cite{Bernardeau} where the non-Gaussian contributions
to the $2$- and $4$-point correlators of temperature were investigated by
expansion in the angular displacement ${\bf g}$, and the results of Ref.\
\cite{Bernardeau} are of course reproduced by the leading-order expressions
given by our Eqs.\ (\ref{C2Ans})--(\ref{C4Ans}). As also noted by Bernardeau
\cite{Bernardeau}, the correction to the power spectrum given by Eq.\ (\ref
{C2Ans}) corresponds, in the lowest order in $C_g$, to a more precise
formula derived in Ref.\ \cite{Seljak96}.

Having treated the simplest case of Gaussian fields, we now consider the possible
generalization of our method to non-Gaussian initial conditions and lensing
source distributions. Such a generalization can be implemented by
introducing new kinds of diagrams contributing to the cumulant expansion. The
non-Gaussian effects may not be limited to even-numbered cumulants in that case.
The generating functional of the lensed temperature can still be obtained
from Eq.\ (\ref{ZExp2}), where one needs to put in a non-Gaussian generating
functional $Z_g\left[ {\bf H}\right] $ to represent the lensing sources as well
as a non-Gaussian term (e.g.\ cubic or quartic in $J$) in the exponential
for each non-vanishing cumulant of the initial CMB distribution.
Then a reasoning similar to the one presented here and in Appendix \ref
{App:GenFunc} will again lead to a diagrammatic representation of cumulants. If
the initial CMB distribution has nonzero cumulants of odd order,
vertices giving odd powers of $J$ will also be present in the diagrams.
On the other hand, non-Gaussianities in the lensing sources lead to many-legged
``propagators'' (Fig.\ 3). For a generic weakly non-Gaussian distribution,
one would expect the magnitude of an $n$-th order cumulant $%
C_g^{\left( n\right) }$ of the lensing sources to be of order $\left(
C_g\right) ^{n/2}$. Then, despite the infinite proliferation of
vertices and propagators, the expression for any given cumulant will
always be finite in all orders in $C_g$ and the leading order term will
usually be given by few diagrams containing low-order vertices and propagators.

Although the signature of lensing becomes less clear-cut for a non-Gaussian
initial CMB distribution, one can still distinguish some intrinsic non-Gaussian
features from those due to lensing. For instance, one can show that the cumulant
$C_n\left( {\bf x},{\bf x},...,{\bf x}\right) $ is unchanged after lensing.

Our formalism seems to require knowledge of the full generating functionals
$Z_0\left[ J\right] $ and $Z_g\left[ {\bf H}\right] $ for the lensing sources
and original CMB distributions, but in fact leading-order results can be obtained
using only the first few cumulants of those distributions, such as the
three-point cumulant of the lensing sources
$C_g^{\left( 3\right) }$ which contributes to $C_4$ according to the diagram
in Fig.\ 3b. The cumulants of the lensing sources can in principle be
estimated from $N$-body cosmological simulations or from galaxy surveys, which is
an interesting direction for future work.

\section{Expected effect in Gaussian CDM models}

\label{Sec:Observable}

Many different statistics have been proposed and used to probe for
non-Gaussian features in the CMB. Here we shall consider a form of the
cumulants-based criterions
suggested in \cite{Cumulants} which in a way generalizes the skewness and
kurtosis \cite{SkewAndKurt} and the genus statistics \cite{Genus}.

Our particular form of the cumulant criterion is related to the morphological
descriptors of Gaussianity based on
Minkowski functionals \cite{MFOthers,MFPaper}, which
are a generalization of the topological genus. Minkowski functionals have
been applied in Ref.\ \cite{MFPaper} to regularly pixelized maps, and it was
found that all three morphological descriptors (the area, the boundary length,
and the genus of temperature excursion sets) are
effectively determined by the three-point distribution of temperature at
nearest neighbor pixels in the map. One might as well inspect this
three-point distribution directly and investigate whether it is Gaussian;
such analysis would be in principle equivalent to inspection of the Minkowski
functional curves calculated from the same map. A standard method of
checking that a distribution is Gaussian is by computing its cumulants \cite
{Kendall}. Thus we arrive to a criterion of Gaussianity of a CMB map which
consists of examination of various cumulants of the three-point distribution
of temperature at neighbor pixels. Note that the evaluation
of a given cumulant requires to compute certain moments of temperature at
neighbor pixel points, and since there is a fixed number of neighbors for
each pixel, the total number of necessary operations is linear in the number
of pixels in the map.

For the purposes of this paper, however, we shall limit ourselves to
considering the two-point cumulants of the distribution of temperature,
computed for pairs of points separated by a fixed angular distance $\theta $%
. The cumulants of a two-variable distribution are labeled by two indices
and may be defined by

\begin{equation}
\chi _{mn}\left( \theta \right) =i^{m+n}\left. \frac{\partial ^m}{\partial
p^m}\frac{\partial ^n}{\partial q^n}\right| _{x,y=0}\ln \tilde{f}_\theta
\left( p,q\right)  \label{Chi2Def}
\end{equation}
where $\tilde{f}_\theta \left( p,q\right) $ is the generating function for
the joint distribution. To be definite, we denote the two field values by $%
f\left( 0\right) $ and $f\left( \theta \right) $, although we imply in our
expressions below that averaging is performed over all pairs of points
separated by $\theta $. The formulae for the practical evaluation of
cumulants from a given map follow from Eq.\ (\ref{Chi2Def}); they are
simplified if the field $f\left( {\bf x}\right) \equiv \Delta T\left( {\bf x}%
\right) /T$ has zero mean, as we shall assume. The cumulant $\chi _{11}$
corresponds to the lensed power spectrum, while  $\chi _{mn}$ with $m+n>2$
represent the non-Gaussian signal.
We shall be particularly interested in the cumulant
\begin{equation}
\chi _{22}\left( \theta \right) =\left\langle f\left( 0\right) ^2f\left(
\theta \right) ^2\right\rangle -2\left\langle f\left( 0\right) f\left(
\theta \right) \right\rangle ^2-\left\langle f\left( 0\right)
^2\right\rangle \left\langle f\left( \theta \right) ^2\right\rangle
\label{Chi22TT}
\end{equation}
which characterizes the (appropriately collapsed) $4$-point correlation function
of temperature.
Since the cumulants defined by Eq.\ (\ref{Chi2Def}) have dimension $f^{m+n}$,
it is convenient to normalize them to units of standard deviation $\sigma
_0^2\equiv \left\langle f\left( 0\right) ^2\right\rangle $,
\begin{equation}
\overline{\chi }_{mn}\left( \theta \right) \equiv \chi _{mn}\left( \theta
\right) \sigma _0^{-\left( m+n\right) }.  \label{ChiBarmn}
\end{equation}

The expected value of the cumulants $\chi _{mn}\left( \theta \right) $ for
the lensed sky is computed in Appendix \ref{App:Chi} where it is shown that
the only nonzero cumulants are those with $m=n$. To characterize the
magnitude of the effect it is convenient to introduce the parameter
\begin{equation}
\sigma _g^2\left( \theta \right) \equiv C_g\left( 0\right)
-C_g\left( \theta \right)
\end{equation}
which describes the relative deviation of photons initially separated by
angular distance $\theta $ (it corresponds to $\sigma ^2\left( \theta
\right) $ of Ref.\ \cite{Seljak96}). Up to $O\left( \sigma _g^6\right) $, the
first few cumulants of the lensed sky are (Eqs.\ (\ref{Chi11A})--(\ref{Chi33A}))
\begin{equation}
\chi _{11}\left( \theta \right) =C_0\left( \theta \right) +\sigma _g^2\left[
C_0^{\prime \prime }+\frac{C_0^{\prime }}\theta \right] +\frac 14\sigma
_g^4\left[ C_0^{\prime \prime \prime \prime }+\frac 2\theta C_0^{\prime
\prime \prime }-\frac 1{\theta ^2}C_0^{\prime \prime }+\frac 1{\theta ^3}%
C_0^{\prime }\right] +O\left( \sigma _g^6\right) ,  \label{Chi11Ans}
\end{equation}
\begin{equation}
\chi _{22}\left( \theta \right) =4\sigma_g^2\left( C_0^{\prime }\right) ^2
+4\sigma _g^4\left[ \left( C_0^{\prime \prime }\right) ^2-\left( \frac{%
C_0^{\prime }}\theta \right) ^2+2C_0^{\prime }\left( C_0^{\prime \prime
\prime }+\frac{C_0^{\prime \prime }}\theta \right) \right] +O\left( \sigma
_g^6\right) ,  \label{Chi22Ans}
\end{equation}
\begin{equation}
\chi _{33}\left( \theta \right) =72\sigma _g^4\left( C_0^{\prime }\right)
^2C_0^{\prime \prime }+O\left( \sigma _g^6\right) .  \label{Chi33Ans}
\end{equation}
(In these expressions we imply that $\sigma _g^2$ and all derivatives of the
correlation function $C_0\left( x\right) $ are evaluated at $x=\theta $.)
Cumulants $\chi _{nn}$ for $n>3$ are of order $O\left( \sigma _g^6\right) $
or smaller, while all other cumulants vanish. This is the specific signature
of the gravitational lensing.

Now we turn to numerical estimates of the non-Gaussian signal measured by $%
\chi _{22}\left( \theta \right) $. Our objective is to find the angular
scale $\theta $ at which the effect is most significant and to determine
whether the signal is within our observational capability. To obtain a
numerical answer, we need to choose the parameters of a particular cosmological
model (the most obvious ones to vary are $h$, $\Omega _b$, $\Omega _\Lambda $
and $n_s$) and compute the two correlators, $C_0\left( \theta \right) $ for
the CMB before lensing and $\sigma _g\left( \theta \right) $ for the lensing
sources. The resulting functions are then substituted into Eqs.\ (\ref{Chi11Ans}%
)--(\ref{Chi33Ans}). The CMB power spectrum $l\left( l+1\right) C_l/\left(
2\pi \right) $ can be generated by the program {\tt CMBFAST} \cite{CMBFAST}
and converted to $C\left( \theta \right) $ using Eq.\ (\ref{CxFromCl}); the
same power spectrum is used to estimate the Gaussian cumulant variances
(cf.\thinspace Eq.\ (\ref{VarChi22})). The latest version of {\tt CMBFAST}
also computes the effect of gravitational lensing on the power spectrum
following Ref.\ \cite{Seljak96}; the author used the {\tt CMBFAST} code to
directly output the lens correlation function $\sigma _g\left( \theta
\right) $.

The resulting dimensionless cumulants
$\bar{\chi}_{22}\left( \theta \right) $ and $\bar{\chi}%
_{33}\left( \theta \right) $ are plotted in Fig.\ 4. We chose two
cosmological models which show the typical range and behavior of the
cumulants. The generic prediction for all models is a peak in $\bar{\chi}%
_{22}$ of magnitude $\sim 10^{-2}$-$10^{-3}$ at $\theta _{*}\sim 10^{\prime
}$-$20^{\prime }$. In open models ($\Omega <1$), the peak is generally lower
and occurs at larger observed angular scales than in flat models. The
numerical results also show that the higher cumulants (as well as the
higher-order terms in $\chi _{22}$) are several orders of magnitude smaller,
as expected. The variance of the cumulant $\bar{\chi}_{22}$ is
typically $\sim 10^{-3}$, which as our graphs show only allows
detection at scales near the angle $\theta =\theta _{*}$ of the maximum
signal. It is necessary to note that the variance of the cumulant $\bar{\chi}%
_{22}$ given by Eq.\ (\ref{VarChi22}) is computed for the theoretical limit
of infinite-resolution, noiseless, full-sky maps, and a real experiment will
give a larger error bar. If the temperature points were completely
uncorrelated and pixels were squares of angular size $\theta _0$, the
distribution of pairs at angle $\theta $ would be estimated from $N\sim 8\pi
\theta _{*}/\theta _0^3$ independent pairs.
For the relative standard error $1/\sqrt{N}$ to be less than $10^{-3}$, the map
resolution must be $\theta _0<15^{\prime }$. This angular scale is beyond the
capabilities of MAP but is at the lower limit of the PLANCK resolution.

It is also easy to see that pixel noise present in the data does not affect
the results as long as the signal-to-noise ratio is somewhat larger than $1$.
Since the cumulants are additive, the observed value of $\chi _{22}$ is
the sum of the cumulant values of signal and noise. Uncorrelated pixel noise
will only contribute to the
variance of $\chi _{22}$ which depends on the fourth power of
the correlation function. Therefore the contribution to the variance
due to pixel noise will be smaller than the intrinsic variance
by the signal-to-noise ratio to the fourth power.

In conclusion, we note that the non-Gaussian effect of gravitational
lensing is a necessary component of the CMB signal in all cosmological
models, and in inflationary scenarios it leads to deviations from the
Gaussian distribution which are small but observable by next-generation
satellite missions. The formalism of generating functionals uncovers the
signature of gravitational lensing, and we have shown that certain
cumulant-based statistics can detect it on subdegree scales.

\section*{Acknowledgments}

The author is grateful to Richard Battye, Neil Cornish, Arthur Hebecker, Uros
Seljak, and Teo Turgut for useful discussions. The author used and adapted
the {\tt CMBFAST} code by Uros Seljak and Matias Zaldarriaga \cite{CMBFAST}.
This work was supported by PPARC rolling grant GR/L21488.

\appendix

\section{Weakly non-Gaussian random fields}
\label{App:WeaklynonG}

Here we describe the formalism of generating functionals we use to
characterize the non-Gaussian features of the CMB. We also give a definition of
cumulants.

Recall that the generating function $\tilde{f}\left( k_1,...,k_n\right) $ of
the joint probability density $p\left( x_1,...,x_n\right) $ for a set of
random variables ${\bf x}\equiv \left\{ x_1,...,x_n\right\} $ is defined by
\begin{equation}
\tilde{f}\left( {\bf k}\right) {\bf \equiv }\left\langle e^{-i{\bf kx}%
}\right\rangle _{{\bf x}}=\int \exp \left( -ik^ix_i\right) p\left( {\bf x}%
\right) d{\bf x}.  \label{FinGenFDef}
\end{equation}
The expectation value of any function $A$ of ${\bf x}$ (treated as a formal
series expansion in ${\bf x}$) can be then obtained by differentiation,
\begin{equation}
\left\langle A\left( {\bf x}\right) \right\rangle =A\left[ i\frac \partial {%
\partial {\bf k}}\right] _{{\bf k}=0}\tilde{f}\left( {\bf k}\right) .
\end{equation}
A description of the distribution of variables $x_i$ in terms of the
generating function $\tilde{f}\left( {\bf k}\right) $ is equivalent to the
description in terms of the joint probability density.

In the infinite-dimensional case of random functions (``fields'') on a
Euclidean space, we consider the generating functional $Z\left[ J\right] $
defined as the average
\begin{equation}
Z\left[ J\right] \equiv \left\langle \exp \left[ -i\int J({\bf x})f({\bf x})d%
{\bf x}\right] \right\rangle _{f({\bf x})}\equiv \left\langle \exp \left[
-iJ_xf_x\right] \right\rangle _f  \label{ZDef}
\end{equation}
over all realizations of $f({\bf x})$. If the generating
functional $Z\left[ J\right] $ for a field $f({\bf x})$ is
given, one can obtain the correlation functions of $f({\bf x})$
and averages of any functional $A$,
\begin{equation}
\left\langle f\left( {\bf x}_1\right) ...f\left( {\bf x}_n\right)
\right\rangle =i^n\left. \frac{\delta ^n}{\delta J\left( {\bf x}_1\right)
...\delta J\left( {\bf x}_1\right) }\right| _{J=0}Z\left[ J\right] ,
\end{equation}
\begin{equation}
\left\langle A\left[ f\left( {\bf x}\right) \right] \right\rangle =A\left[ i%
\frac \delta {\delta J\left( {\bf x}\right) }\right] _{J=0}Z\left[ J\right] .
\label{ExpectA}
\end{equation}
This implies the normalization $\left\langle 1\right\rangle =Z\left[
0\right] =1$.

Let us give an example of how the generating functional $Z\left[ J\right] $
of a random field $f({\bf x})$ can be directly used to evaluate
some field statistics, such as the ones used for CMB analysis. If one is
interested in finding the distribution of values $f({\bf x})$
at certain given points ${\bf x}_1$, ..., ${\bf x}_n$, one can obtain such
distributions directly from $Z\left[ J\right] $. Again, we should look for
the generating function $\tilde{f}\left( k_1,...,k_n\right) $ for the
distribution of values $f({\bf x}_1)$, ..., $f({\bf x}_n)$,
as defined by Eq.\ (\ref{FinGenFDef}), rather than for the
probability density itself. It is easy to show that the generating function $%
\tilde{f}\left( k_1,...,k_n\right) $ is found by substituting into $Z\left[
J\right] $ a suitable function $J({\bf x})$:
\begin{equation}
\tilde{f}\left( k_1,...,k_n\right) =Z\left[J({\bf x})=k_1\delta \left(
{\bf x}-{\bf x}%
_1\right) +...+k_n\delta \left( {\bf x}-{\bf x}_n\right) \right] .
\end{equation}
Another example: The generating function $\tilde{f}\left( k_0,{\bf k}\right)
$ for the joint distribution of $f$ and its gradient ${\bf \nabla }f$ taken
at a reference point ${\bf x}_0$ is
\begin{equation}
\tilde{f}\left( k_0,{\bf k}\right) =Z\left[ k_0\delta \left( {\bf x}-{\bf x}%
_0\right) -{\bf k\cdot \nabla }\delta \left( {\bf x}-{\bf x}_0\right)
\right] .  \label{FDF}
\end{equation}

We turn now to the question of characterizing non-Gaussian features of
random fields. It is straightforward to show that the generating functional
of a Gaussian distributed random field is a Gaussian functional,
\begin{equation}
Z_G\left[ J\right] =\exp \left[ -i\int \mu ({\bf x})J({\bf x})d{\bf x}-\frac
12\int J({\bf x})J({\bf y})C( {\bf x},{\bf y}) d{\bf x}d{\bf y}%
\right] \equiv \exp \left[ -i\mu J-\frac 12J_xC^{xy}J_y\right] , \label{ZG}
\end{equation}
where $\mu $ is the mean value and $C$ is the correlation function. If the
field in question is homogeneous and isotropic, its mean value is a constant
$\mu \left( {\bf x}\right) \equiv \mu $ and the correlation function depends
only on the distance, $C\left( {\bf x},{\bf y}\right) =C\left( \left| {\bf x}%
-{\bf y}\right| \right) $. A general non-Gaussian field will not have the
generating functional $Z\left[ J\right] $ of the form of Eq.\ (\ref{ZG});
however, we could
attempt to express $Z\left[ J\right] $ as an exponential of a formal series
in $J$, written in condensed notation as
\begin{equation}
Z\left[ J\right] =\exp \left[ -i\mu J-\frac 12J_xJ_yC^{xy}+\frac{\left(
-i\right) ^3}{3!}J_xJ_yJ_zC_3^{xyz}+...\right] ,  \label{ZGen}
\end{equation}
with some functions $C_n\left( {\bf x}_1,...,{\bf x}_n\right) $, $n=3,4,...$
as parameters. For this representation to be possible, it is necessary for
all moments of the distribution to be finite, which of course is not the
case in general. However, we are interested in distributions which are
``weakly non-Gaussian'', and in particular, we expect all moments of such
distributions to exist. So we shall take the possibility of a series
representation (Eq.\ (\ref{ZGen})) for the generating functional of
a random field as a definition of its {\em weak} non-Gaussianity.

We regard the coefficients $C_n\left( {\bf x}_1,...,{\bf x}_n\right) $, $n=3$,
$4$, ... (which are in fact the {\em cumulants} of the distribution of $f$)
as general descriptors of the non-Gaussian features of the random field $f$.
The cumulants can be related to the correlation functions by rewriting
Eq.\ (\ref{ZGen}) as
\begin{equation}
C_n\left( {\bf x}_1,...,{\bf x}_n\right) =i^n\left. \frac{\delta ^n}{\delta
J\left( {\bf x}_1\right) ...\delta J\left( {\bf x}_1\right) }\right|
_{J=0}\ln Z\left[ J\right]   \label{CThruLnZ}
\end{equation}
and replacing the resulting derivatives of $Z\left[ J\right] $ at $J=0$ by
the corresponding correlation functions. It follows that
$C_1=\mu $ and $C_2\left( {\bf x},{\bf y}\right) =C\left( {\bf x}-{\bf y}%
\right) $ are the mean and the correlation function of the distribution, and
for example the third-order cumulant is
\begin{equation}
C_3\left( {\bf x},{\bf y},{\bf z}\right) =\left\langle f\left( {\bf x}%
\right) f\left( {\bf y}\right) f\left( {\bf z}\right) \right\rangle -\left[
C\left( {\bf x}-{\bf y}\right) +C\left( {\bf y}-{\bf z}\right) +C\left( {\bf %
z}-{\bf x}\right) \right] \mu +2\mu ^3.
\end{equation}
Because $\ln Z\left[ J\right] $ is the generating functional of cumulants,
similarly to the generating functional of connected Green's functions in
quantum field theory, the cumulants are sometimes called the ``connected
moments''. For a Gaussian distribution, all cumulants $C_n$ with $n\geq 3$
vanish by construction.

A convenient property of the cumulants $C_n$ (to which they owe their name)
is additivity: if $f$ is a sum of two statistically independent
distributions $g$ and $h$, then the generating functional $Z_f$ for $f$ will
be the product $Z_f\left[ J\right] =Z_g\left[ J\right] Z_h\left[ J\right] $
of the generating functionals for $g$ and $h$. Because of the logarithm in
Eq.\ (\ref{CThruLnZ}), the cumulants for $g$ and $h$ will add to give the
cumulants for $f$.

The description of a non-Gaussian field in terms of the cumulants $C_n$
is a compact way to put together the
information from all $n$-point correlation functions.

\section{Generating functional for composition of fields}

\label{App:GenFunc}

Here we compute the generating functional for the lensed field of Eq.\ (\ref
{LensedField}). The average over the space of realizations of $T( {\bf x})$
means the average over $T_0\left( {\bf x}\right) $ and ${\bf g}%
( {\bf x}) $. We assume that both $T_0( {\bf x}) $ and
${\bf g}( {\bf x}) $ are Gaussian distributed with generating
functionals
\begin{equation}
Z_0\left[ R\right] =\exp \left[ -\frac 12\int R\left( {\bf x}\right)
C_0\left( {\bf x}-{\bf y}\right) R\left( {\bf y}\right) d{\bf x}d{\bf y}%
\right] \equiv \exp \left[ -\frac 12R_xC_0^{xy}R_y\right] ,  \label{ZG0}
\end{equation}
\begin{equation}
Z_g\left[ {\bf H}\right] =\exp \left[ -\frac 12\int {\bf H}^T( {\bf x}%
) {\bf C}_g\left( {\bf x}-{\bf y}\right) {\bf H}( {\bf y}) d%
{\bf x}d{\bf y}\right] \equiv \exp \left[ -\frac 12{\bf H}_x^T{\bf C}_g^{xy}%
{\bf H}_y\right]   \label{ZGamma}
\end{equation}
(in the last expression, ${\bf H}$ is an $R^n$-valued function and matrix
multiplication is implied between ${\bf C}_g$ and ${\bf H}$).

According to Eq.\ (\ref{ExpectA}), the generating functional for $T\left(
{\bf x}\right) $ is formally written by substituting $i\delta /\delta R$
instead of $T_0$ and $i\delta /\delta {\bf H}$ instead of ${\bf g}$ in Eq.\ (%
\ref{ZDef}),
\begin{eqnarray}
Z\left[ J\right]  &\equiv &\left\langle \exp \left( -i\int J\left( {\bf x}%
\right) T_0\left( {\bf x}+{\bf g}\left( {\bf x}\right) \right) d{\bf x}%
\right) \right\rangle _{T_0,{\bf g}}  \nonumber \\
&=&\exp \left( \int d{\bf x}\,J\left( {\bf x}\right) \frac \delta {\delta
R\left( {\bf x}+i\frac \delta {\delta {\bf H}\left( {\bf x}\right) }\right) }%
\right) _{R=0,\,{\bf H}=0}Z_0\left[ R\right] Z_g\left[ {\bf H}\right] .
\end{eqnarray}

To make the calculation clearer, we first evaluate the average only with
respect to $T_0$, which corresponds to taking the derivative $\delta /\delta
R$.
We temporarily introduce the inverse shift function ${\bf \tilde{g}}\left(
{\bf x}\right) $ which satisfies ${\bf g}\left( {\bf x}\right) ={\bf \tilde{g%
}}\left[ {\bf x}+{\bf g}\left( {\bf x}\right) \right] $ so that we can
change variables
\begin{equation}
\int J\left( {\bf x}\right) T_0\left( {\bf x}+{\bf g}\left( {\bf x}\right)
\right) d{\bf x}=\int J\left( {\bf z}-{\bf \tilde{g}}\left( {\bf z}\right)
\right) T_0\left( {\bf z}\right) d{\bf z}.
\end{equation}
(We shall not need an explicit expression for ${\bf \tilde{g}}\left( {\bf x}
\right) $ in what follows.) Then we obtain
\[
\left\langle \exp \left( -i\int J\left( {\bf x}\right) T_0\left( {\bf x}+%
{\bf g}\left( {\bf x}\right) \right) d{\bf x}\right) \right\rangle
_{T_0}
=Z_0\left[ J\left( {\bf x}-{\bf \tilde{g}}\left({\bf x}\right) \right) \right]
\]
\[
=\exp \left[ -\frac 12\int d{\bf x}\int d{\bf y}\,J\left( {\bf x}-{\bf
\tilde{g}}\left( {\bf x}\right) \right) C_0\left( {\bf x}-{\bf y}\right)
J\left( {\bf y}-{\bf \tilde{g}}\left( {\bf y}\right) \right) \right]
\]
\begin{equation}
=\exp \left[ -\frac 12\int d{\bf x}\int d{\bf y}\,J\left( {\bf x}\right)
C_0\left( {\bf x}-{\bf y}+{\bf g}\left( {\bf x}\right) -{\bf g}\left( {\bf y}%
\right) \right) J\left( {\bf y}\right) \right] \equiv \exp \left[ -\frac 12%
J_x\tilde{C}^{xy}\left[ {\bf g}\right] J_y\right] .
\end{equation}
The correlation function can be expanded as

\begin{eqnarray}
\tilde{C}^{xy}\left[ {\bf g}\right] &\equiv &C_0\left[ {\bf x}-{\bf y}+{\bf g%
}\left( {\bf x}\right) -{\bf g}\left( {\bf y}\right) \right]
=C_0^{xy}+\left[ g^k\left( {\bf x}\right) -g^k\left( {\bf y}\right) \right]
\nabla _kC_0\left( {\bf x}-{\bf y}\right)  \nonumber \\
&&+\frac{\left[ g^k\left( {\bf x}\right) -g^k\left( {\bf y}\right) \right]
\left[ g^l\left( {\bf x}\right) -g^l\left( {\bf y}\right) \right] }2\nabla
_k\nabla _lC_0\left( {\bf x}-{\bf y}\right) +O\left( {\bf g}^3\right) .
\label{CorrExp}
\end{eqnarray}
We now only need to substitute $i\delta /\delta {\bf H}$ instead of ${\bf g}$
to express the generating functional as
\begin{equation}
Z\left[ J\right] =\exp \left[ -\frac 12J_xC_0\left( {\bf x}-{\bf y}+i\frac
\delta {\delta {\bf H}\left( {\bf x}\right) }-i\frac \delta {\delta {\bf H}%
\left( {\bf y}\right) }\right) J_y\right] _{H=0}\exp \left[ -\frac 12{\bf H}%
_x^T{\bf C}_g^{xy}{\bf H}_y\right] .  \label{Ans1}
\end{equation}

The expression of Eq.\ (\ref{Ans1}) is of course understood in the sense of
perturbative expansion as shown by Eq.\ (\ref{CorrExp}), yielding a sum
of terms with functional derivatives of increasing order. In the
context of weak lensing, the correlation function $C_g$ can be regarded as a
small parameter in the sense of Eq.\ (\ref{CGSmall}), and expansion in
powers of $C_g$ is a natural choice. Each pair of functional
differentiations $\delta /\delta {\bf H}$ applied to the Gaussian generating
functional $Z_g\left[ {\bf H}\right] $ would add at least one more power of $%
C_g$, therefore the expansion in $C_g$ is consistent with truncating the
operator series at a finite order of functional derivatives.

Exploring the quantum field theory analogy further, we regard Eq.\ (\ref
{Ans1}) as the generating functional of Green's functions for an interacting
quantum field, with an ``interaction vertex'' for each term of the expansion
in gradients of $C_0$. The connected Green's functions of this theory are
generated by $\ln Z\left[ J\right] $, which in our context is the generating
functional of cumulants. According to standard rules, we can express $\ln
Z\left[ J\right] $ as the sum of all connected diagrams made up of vertices $%
V^{\left( n\right) }$ with any number of legs $n\geq 1$; the propagator is
the matrix ${\bf C}_g\left( {\bf x}-{\bf y}\right) $. Since all functional
derivatives are taken at ${\bf H}=0$, the resulting diagrams are all
``vacuum diagrams'' without external legs (Figs.\ 1, 2).

To obtain the general expression $V^{\left( n\right) }$ for the vertex with $%
n$ legs, we need to consider the $n$-th term in the expansion of the
correlation function,
\begin{eqnarray}
C_0&&\left( {\bf x}-{\bf y}+{\bf g}\left( {\bf x}\right) -{\bf g}\left( {\bf %
y}\right) \right) =C_0\left( {\bf x}-{\bf y}\right) +\left( g^k\left( {\bf x}%
\right) -g^k\left( {\bf y}\right) \right) \nabla _kC_0\left( {\bf x}-{\bf y}%
\right) +...  \nonumber \\
&&+\frac 1{n!}\int \nabla _{k_1}...\nabla _{k_n}C_0\left( {\bf x}-{\bf y}%
\right) \prod_i\left[ \delta \left( {\bf x}_i-{\bf x}\right) -\delta \left(
{\bf x}_i-{\bf y}\right) \right] \prod_ig^{k_i}\left( {\bf x}_i\right) d{\bf %
x}_i+...
\end{eqnarray}
The required vertex coefficient is the factor at $\prod_ig^{k_i}\left( {\bf x%
}_i\right) d{\bf x}_i$ without the $n!$ denominator, integrated with
$d{\bf x}d{\bf y} J({\bf x}) J( {\bf y})/2$,
\begin{equation}
V_{k_1...k_n}^{\left( n\right) }\left( {\bf x}_1,...,{\bf x}_n\right) =\frac 12
\int d{\bf x}d{\bf y}J( {\bf x}) J( {\bf y}) \nabla
_{k_1}...\nabla _{k_n}C_0\left( {\bf x}-{\bf y}\right) \prod_i\left[ \delta
( {\bf x}_i-{\bf x}) -\delta ( {\bf x}_i-{\bf y})
\right] .  \label{VertexN}
\end{equation}
Each leg of a vertex is labeled by a spatial variable ${\bf x}_i$ and an index $k_i$.
The first two vertices computed from Eq.\ (\ref{VertexN}) are
\begin{eqnarray}
V_k^{\left( 1\right) }\left( {\bf x}\right) &=&J({\bf x}) \int
J({\bf y}^{\prime }) \nabla _kC_0\left( {\bf x}-{\bf y}^{\prime
}\right) d{\bf y}^{\prime },  \label{ADef} \\
V_{kl}^{\left( 2\right) }\left( {\bf x},{\bf y}\right) &=&-J({\bf x}%
) J( {\bf y}) \nabla _k\nabla _lC_0\left( {\bf x}-{\bf y}%
\right) +\delta ( {\bf x}-{\bf y}) J( {\bf x}) \int
J( {\bf y}^{\prime }) \nabla _k\nabla _lC_0\left( {\bf x}-{\bf y}%
^{\prime }\right) d{\bf y}^{\prime }  \label{BDef}
\end{eqnarray}
(in the last expressions all integrations are made explicit, and we made use
of the symmetry of the correlation function $C_0$). Summation over indices
and integration over spatial variables are performed when connecting lines,
as usual.

Since we are expanding in $C_g$, the order of the diagram is equal to the
number of the propagators. One can also obtain all diagrams of a given order
by selecting the term of the given order in $C_g$ out of the expansion of $%
Z_g\left[ {\bf H}\right] $ in Eq.\ (\ref{Ans1}) and acting with the
functional derivatives on that term. Since each $n$-th power of $C_g$ is
accompanied by the $(2n)$-th power of ${\bf H}$, the only surviving
derivatives will be of order not greater than $2n$, which limits the
possible vertices in the diagrams to a finite set.

Since each vertex is quadratic in $J$, a connected diagram with $n$ vertices
$V^{\left( k_1\right) }$, ..., $V^{\left( k_n\right) }$ is of order $\left(
C_g\right) ^{k_1+...+k_n}$ and contributes to the cumulant $C_{2n}$. The
lowest-order diagrams are those with one propagator (Fig.\ 1b); they yield
leading-order expressions for $C_2\equiv C$ and $C_4$. Computing the
diagrams in Fig.\ 1b only requires the vertices $V^{\left( 1\right) }$ and $%
V^{\left( 2\right) }$ and gives Eqs.\ (\ref{C2Ans})--(\ref{C4Ans}) after
accounting for the factorial coefficients in Eq.\ (\ref{ZGen})
and combinatorial factors.

\section{Two-point cumulants of the lensed sky}

\label{App:Chi}

The two-point cumulants $\chi _{mn}$ can be straightforwardly obtained from
the generating functional $\ln Z\left[ J\right] $ of Eq.\ (\ref{LnZAns});
as described in Appendix \ref{App:WeaklynonG}, one needs to
substitute the indicator function
\begin{equation}
J_{pq}\left( {\bf x}\right) =p\delta \left( {\bf x}\right) +q\delta \left( {\bf x}%
-\theta \right)  \label{Chi22Ind}
\end{equation}
into the generating functional and collect
terms of the form $p^mq^n$. However, it is somewhat simpler to use Eq.\ (\ref
{Chi22Ind}) directly with Eq.\ (\ref{ZExp1}) to derive the diagram rules
specifically for the generating function $Z_\chi \left( p,q\right) $ of
two-point cumulants,
\begin{equation}
Z_\chi \left( p,q\right) =\left\langle \exp \left[ -\frac{\sigma _0^2}2%
\left( p^2+q^2\right) -pqC_0\left[ \theta +{\bf g}\left( \theta \right) -%
{\bf g}\left( 0\right) \right] \right] \right\rangle _{{\bf g}}.
\label{ZChi}
\end{equation}
Since only the difference $\Delta {\bf g}\equiv {\bf g}\left( \theta \right)
-{\bf g}\left( 0\right) $ enters Eq.\ (\ref{ZChi}), we can simplify it by
considering $\Delta {\bf g}$ as a random (vector-valued) variable whose
distribution is specified by the generating function
\begin{eqnarray}
Z_{\Delta g}\left( {\bf h}\right) &\equiv &\left\langle \exp \left[ -i{\bf %
h\cdot }\left( {\bf g}\left( \theta \right) -{\bf g}\left( 0\right) \right)
\right] \right\rangle  \nonumber \\
&=&Z_g\left[ {\bf H}=\left( \delta \left( x-\theta \right) -\delta \left(
x\right) \right) {\bf h}\right] =\exp \left[ -h^2\left( C_g\left( 0\right)
-C_g\left( \theta \right) \right) \right] .
\end{eqnarray}
By a procedure similar to that we used above, we arrive to the following
diagram rules. For each $n=1$, $2$, ... there is a vertex $V_n$ with $n$
legs labeled by indices $k_1$ to $k_n$, and the coefficient corresponding to
the vertex is
\begin{equation}
V_n=pq\nabla _{k_1}...\nabla _{k_n}C_0\left( \theta \right) ,
\end{equation}
where the gradients are taken in the $2$-dimensional Euclidean space, for
instance
\begin{eqnarray}
\nabla _kC_0\left( x\right) &=&C_0^{\prime }\left( x\right) \frac{x_k}{%
\left| {\bf x}\right| }, \\
\nabla _k\nabla _lC_0\left( x\right) &=&C_0^{\prime \prime }\left( x\right)
\frac{x_kx_l}{\left| {\bf x}\right| ^2}+C_0^{\prime }\left( x\right) \frac{%
x^2\delta _{kl}-x_kx_l}{\left| {\bf x}\right| ^3} .
\end{eqnarray}
(Since we only consider small patches of the sky, we use
Cartesian $2$-dimensional vectors ${\bf x}$ interchangeably with the angular
coordinate $\theta $ to denote points.) The ``propagator'' is
\begin{equation}
2\left[ C_g\left( 0\right) -C_g\left( \theta \right) \right] \delta
_{kl}\equiv 2\sigma _g^2\left( \theta \right) \delta _{kl}
\label{Propagator2}
\end{equation}
where the $\delta $-tensor effectively sums over indices labeling the vertex
legs. As usual in perturbation theory, one should count all diagrams
differing by the way the vertices are connected and divide by the resulting
combinatorial factor.

The cumulant $\chi _{mn}$ is obtained by adding all diagrams which give
$p^mq^n$ (with appropriate combinatorial factors) and multiplying by
$(m!n!)$. Since $p$ and $q$ enter the diagrams only in the
combination $pq$, it follows that only cumulants $\chi _{mn}$ with $m=n$
are nonzero. The cumulant $\chi _{11}$
is given by the sum of diagrams with one vertex;
the lowest-order ones are the first diagram in Fig.\ 1(b) and the diagram in
Fig.\ 2(a). These two diagrams give
\begin{eqnarray}
\chi _{11}\left( \theta \right) &=&C_0\left( \theta \right) +\sigma
_g^2\left[ C_0^{\prime \prime }+\frac{C_0^{\prime }}\theta \right]  \nonumber
\\
&&+\frac 14\sigma _g^4\left[ C_0^{\prime \prime \prime \prime }+\frac 2\theta
C_0^{\prime \prime \prime }-\frac 1{\theta ^2}C_0^{\prime \prime }+\frac 1{%
\theta ^3}C_0^{\prime }\right] +O\left( \sigma _g^6\right) .  \label{Chi11A}
\end{eqnarray}
(Here and below we imply that $\sigma _g^2$ and all derivatives of the
correlation function $C_0\left( x\right) $ are evaluated at $x=\theta $.)
Diagrams with two vertices in Fig.\ 1(c) and Fig.\ 2(b) contribute to the
cumulant $\chi _{22}\left( \theta \right) $,
\begin{equation}
\chi _{22}\left( \theta \right) =4\sigma_g^2\left( C_0^{\prime }\right) ^2
+4\sigma _g^4\left[ \left( C_0^{\prime \prime }\right) ^2-\left( \frac{%
C_0^{\prime }}\theta \right) ^2+2C_0^{\prime }\left( C_0^{\prime \prime
\prime }+\frac{C_0^{\prime \prime }}\theta \right) \right] +O\left( \sigma
_g^6\right) .  \label{Chi22A}
\end{equation}
The diagram in Fig.\ 2(c) gives the leading-order expression for the cumulant
$\chi _{33}$,
\begin{equation}
\chi _{33}\left( \theta \right) =72\sigma _g^4\left( C_0^{\prime }\right)
^2C_0^{\prime \prime }+O\left( \sigma _g^6\right) .  \label{Chi33A}
\end{equation}

Finally, estimate the variance of the cumulant
$\chi _{22}\left( \theta \right) $ for
Gaussian skies. Because the temperature values are correlated, the set of all
pairs of points separated by the angular distance $\theta $ is not a
statistically independent sample. The cumulant estimator $\hat{\chi}_{22}$
for a random function $f( {\bf x})$ is a random variable expressed through
$f( {\bf x})$ as  (cf.\ Eq.\ (\ref{Chi22TT}))
\begin{equation}
\hat{\chi}_{22}=\frac 1A\int_{\left| {\bf x}-{\bf x}^{\prime }\right|
=\theta }f^2( {\bf x}) f^2( {\bf x}^{\prime }) d{\bf x}%
d{\bf x}^{\prime }-2\left[ \frac 1A\int_{\left| {\bf x}-{\bf x}^{\prime
}\right| =\theta }f( {\bf x}) f( {\bf x}^{\prime }) d%
{\bf x}d{\bf x}^{\prime }\right] ^2-\left[ \frac 1A\int f^2( {\bf x}%
) d{\bf x}\right] ^2
\end{equation}
(here $A$ is the area of the integration region).
The expectation value of $\left( \hat{%
\chi}_{22}\right) ^2$ involves at least a double integration over the observed region,
\begin{equation}
\left\langle \left( \hat{\chi}_{22}\right) ^2\right\rangle =\frac 1{A^2}%
\int_{\left| {\bf x}-{\bf x}^{\prime }\right| =\theta }d{\bf x}d{\bf x}%
^{\prime }\int_{\left| {\bf y}-{\bf y}^{\prime }\right| =\theta }d{\bf y}d%
{\bf y}^{\prime }f^2( {\bf x}) f^2( {\bf x}^{\prime })
f^2( {\bf y}) f^2( {\bf y}^{\prime }) +...
\label{Chi22DI}
\end{equation}
The angle $\theta $ is typically much smaller than the separation between $%
{\bf x}$ and ${\bf y}$ in the integrals of Eq.\ (\ref{Chi22DI}), and we can
replace ${\bf x}^{\prime }$ by ${\bf x}$, ${\bf y}^{\prime }$ by ${\bf y}$
and so on in that equation:
\begin{eqnarray}
\left\langle \left( \hat{\chi}_{22}\right) ^2\right\rangle &\approx &\frac 1{%
A^2}\int f^4( {\bf x}) f^4( {\bf y}) d{\bf x}d{\bf y}-%
\frac 6{A^3}\int f^4( {\bf x}) f^2( {\bf y}) f^2(
{\bf z}) d{\bf x}d{\bf y}d{\bf z} \\
&&+\frac 9{A^4}\int f^2( {\bf w}) f^2( {\bf x})
f^2( {\bf y}) f^2( {\bf z}) d{\bf w}d{\bf x}d{\bf y}d%
{\bf z},  \label{Chi22SI}
\end{eqnarray}
where the integrations in ${\bf w}$, ${\bf x}$, ${\bf y}$, ${\bf z}$ are
over all sky. This approximation is equivalent to replacing $\chi
_{22}\left( \theta \right) $ by the one-point cumulant
\begin{equation}
\chi _4=\left\langle f^4\right\rangle -3\left\langle f^2\right\rangle ^2
\end{equation}
which is adequate for estimating the approximately $\theta $-independent
variance of $\chi _{22}\left( \theta \right) $ at small $\theta$. If the distribution of $f$
is Gaussian with zero mean and correlation function $C_0\left( {\bf x}%
\right) $, then Eq.\ (\ref{Chi22SI}) gives after some algebra
\begin{eqnarray}
\left\langle \left( \hat{\chi}_{22}\right) ^2\right\rangle &\approx &\frac{%
432}{A^3}\int C_0\left( {\bf x}\right) C_0\left( {\bf y}\right) C_0\left(
{\bf z}-{\bf x}\right) C_0\left( {\bf z}-{\bf y}\right) d{\bf x}d{\bf y}d%
{\bf z}  \nonumber \\
&&-\frac{36}{A^2}\left[ \int C_0^2\left( {\bf x}\right) d{\bf x}\right] ^2+%
\frac{24}A\int C_0^4\left( {\bf x}\right) d{\bf x}.
\end{eqnarray}
The first two integrals can be expressed through the angular power spectrum,
and we obtain
\begin{equation}
\left\langle \left( \hat{\chi}_{22}\right) ^2\right\rangle \approx \frac{432%
}{\left( 4\pi \right) ^4}\sum_l\left( 2l+1\right) C_l^4-\frac{36}{\left(
4\pi \right) ^2}\left[ \sum_l\left( 2l+1\right) C_l^2\right] ^2+12\int_0^\pi
C_0^4\left( \theta \right) \sin \theta d\theta .  \label{VarChi22}
\end{equation}
This is the final formula we use for the variance of the cumulant estimator $%
\chi _{22}\left( \theta \right) $ at small angles $\theta $.

\section{An algorithm for computing $n$-point cumulants}

Cumulants of a distribution of $n$ variables $x_1$, ..., $x_n$ are
labeled by $n$ indices $k_1$, ..., $k_n$ and can be defined by
\begin{equation}
\chi _{k_1...k_n}\equiv \left. \left( i\frac \partial {\partial p_1}\right)
^{k_1}...\left( i\frac \partial {\partial p_n}\right) ^{k_n}\right|
_{x_i=0}\ln \tilde{f}\left( p_1,...,p_n\right) ,  \label{AllChis}
\end{equation}
where $\tilde{f}$ is the generating function for the joint distribution of
$x_i$. We call $k_1+...+k_n$ the {\em order} of the cumulant $\chi
_{k_1...k_n}$. The moments of the distribution are expressed through the
same generating function as
\begin{equation}
\mu _{k_1...k_n}\equiv \left\langle x_1^{k_1}...x_n^{k_n}\right\rangle
=\left. \left( i\frac \partial {\partial p_1}\right) ^{k_1}...\left( i\frac
\partial {\partial p_n}\right) ^{k_n}\right| _{x_i=0}\tilde{f}\left(
p_1,...,p_n\right) .  \label{AllMoments}
\end{equation}
Although Eqs.\ (\ref{AllChis})--(\ref{AllMoments}) allow to express all
cumulants through the moments, a direct calculation for all combinations of
indices is cumbersome to implement. We present a compact recursive algorithm
for computing all $\chi $'s whereby the cumulants corresponding to all
combinations of $k$'s are evaluated using the previously computed cumulants
of lower orders.

For the case $n=1$, this algorithm is given in Ref.\ \cite{Kendall} and is
based on the identity
\begin{equation}
\mu _s=\sum_{k=0}^s{s-1 \choose k-1}\chi _k\mu _{s-k}.  \label{MuId1}
\end{equation}
Here we imply that $\chi _0\equiv 0$ and $\mu _0=1$,
and that binomial coefficients with negative entries vanish. From Eq.\
(\ref{MuId1}) we can recursively express $\chi _s$ through the lower-order
cumulants,
\begin{equation}
\chi _s=\mu _s-\sum_{k=0}^{s-1}%
{s-1 \choose k-1}
\chi _k\mu _{s-k}.  \label{ChiRec1}
\end{equation}

In the general case $n>1$, we need an analog of the identity of Eq.\ (\ref
{MuId1}). The resulting formula turns out to be
\begin{equation}
\mu _{s_1...s_n}=\sum_{k_1=0}^{s_1}...\sum_{k_n=0}^{s_n}\frac{k_1+...+k_n}{%
s_1+...+s_n}{s_1 \choose k_1}...{s_n \choose k_n}
\chi _{k_1...k_n}\mu _{s_1-k_1,...,s_n-k_n},  \label{MuIdN}
\end{equation}
from which one obtains the cumulant $\chi _{s_1...s_n}$ as a function of
moments $\mu $ and cumulants $\chi $ of lower orders. Using Eq.\ (\ref{MuIdN}%
), one can organize the computation of all cumulants order by order in an
iterative manner. Given $N$ sample realizations of the set of random
variables $\left\{ x_i\right\} $, one first calculates the moments $\mu
_{s_1...s_n}$ up to the required order; this takes a number of operations
linear in $N$. The first-order cumulants $\chi _{100...0}$, $\chi _{010...0}$
etc.\ are equal to the similarly labeled moments. Thus having obtained the
base of induction, one iterates through increasing order of the cumulants index
by index, using Eq.\ (\ref{MuIdN}) to express the next cumulant
through moments and previously computed cumulants of lower orders.

Now we outline the derivation of Eq.\ (\ref{MuIdN}). We combine Eqs.\ (\ref
{AllChis}) and (\ref{AllMoments}) into an identity
\begin{equation}
\exp \left[ \sum_{k_i=0}^\infty \chi _{k_1...k_n}\frac{p_1^{k_1}...p_n^{k_n}%
}{k_1!...k_n!}\right] =\sum_{k_i=0}^\infty \mu _{k_1...k_n}\frac{%
p_1^{k_1}...p_n^{k_n}}{k_1!...k_n!}\equiv \sum_{\left[ k\right] }\mu
_{\left[ k\right] }\frac{\left[ p^k\right] }{\left[ k\right] !}.
\label{ExpId}
\end{equation}
Regarding $\mu _{k_1...k_n}\equiv \mu _{\left[ k\right] }$ as functions
of $\chi $'s, we take the partial derivative of Eq.\ (\ref{ExpId}) with
respect to $\chi _{\left[ k\right] }$ for a fixed multi-index $\left[
k\right] $. By equating the coefficients at $p_1^{s_1}...p_n^{s_n}\equiv
\left[ p^s\right] $ we obtain the relation
\begin{equation}
\frac \partial {\partial \chi _{\left[ k\right] }}\mu _{\left[ s\right] }=%
\frac{\left[ s\right] !}{\left[ k\right] !\left[ s-k\right] !}\mu _{\left[
s-k\right] },  \label{DiffIdN}
\end{equation}
where the multi-index $\left[ s-k\right] $ means $\left\{ \left(
s_1-k\right) ,...,\left( s_n-k_n\right) \right\} $ in our condensed
notation. We also introduce for convenience the generalized binomial
coefficient
\begin{equation}
{\left[ s\right]  \choose \left[ k\right] }
\equiv \frac{\left[ s\right] !}{\left[ k\right] !\left[ s-k\right] !}=
{s_1 \choose k_1}...{s_n \choose k_n}.  \label{GenBinomial}
\end{equation}

The form of Eq.\ (\ref{MuId1}) leads us to conjecture that the generalized
identity should look like
\begin{equation}
\mu _{\left[ s\right] }=\sum_{k_1=0}^{s_1}...\sum_{k_n=0}^{s_n}%
{s \atopwithdelims\{\} k}
\chi _{\left[ k\right] }\mu _{\left[ s-k\right] }  \label{MuIdConj}
\end{equation}
with some combinatorial coefficients ${s \atopwithdelims\{\} k}$
which should be similar to those of Eq.\ (\ref{GenBinomial}). The
requirement that these coefficients must satisfy Eq.\ (\ref{DiffIdN}) leads
after some algebra to
\begin{equation}
{s \atopwithdelims\{\} k}=\frac{\sum k_i}{\sum s_i}
{\left[ s\right]  \choose \left[ k\right] },
\end{equation}
which yields Eq.\ (\ref{MuIdN}) after substitution into Eq.\ (\ref{MuIdConj}%
).


\begin{figure}[tbh]
\epsfysize 4cm \epsffile{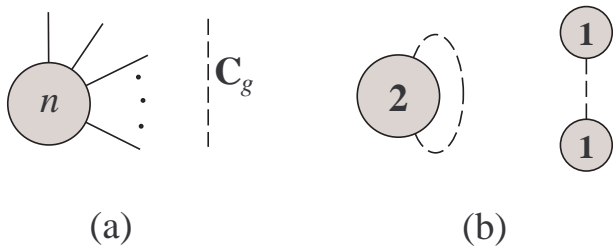}
 \label{Fig:LowestDiags}
\caption{ Diagram representation of the generating function for the lensed
sky. (a) A vertex labeled $n$ has $n$ legs, corresponding to $n$-th
derivative of $C_0$. The propagator corresponds to the lensing correlation
function ${\bf C}_g$. (b) Lowest-order diagrams with one propagator lead to
corrections in the power spectrum and a non-Gaussian $4$-point correlation
function.}

\end{figure}

\begin{figure}[tbh]
\epsfysize 4cm \epsffile{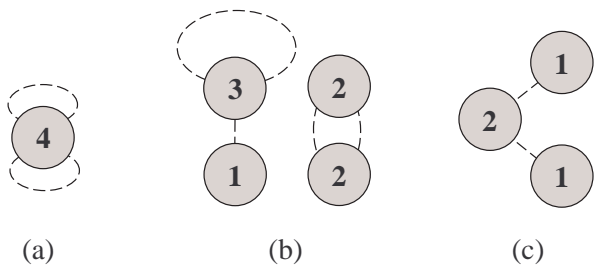}
 \label{Fig:NextDiags}
\caption{ Diagrams leading to second-order effects: (a) correction to the
power spectrum, (b) second-order terms in the cumulant $\chi_{22}$, (c)
leading-order term in the cumulant $\chi_{33}$. }
\end{figure}

\begin{figure}[tbh]
\epsfysize 4cm \epsffile{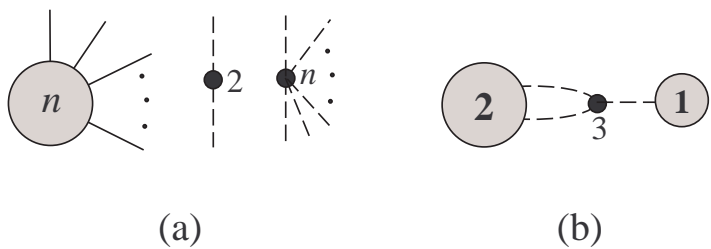}
 \label{Fig:NonGDiags}
\caption{ Diagrams for non-Gaussian initial conditions. (a) In addition to
previous types of diagrams, there are ``propagators'' with more than two legs which
correspond to non-Gaussian lensing sources. (b) An example of a new diagram
contributing to the $4$-point cumulant.}
\end{figure}

\begin{figure}[tbh]
\epsfysize 16cm \epsffile{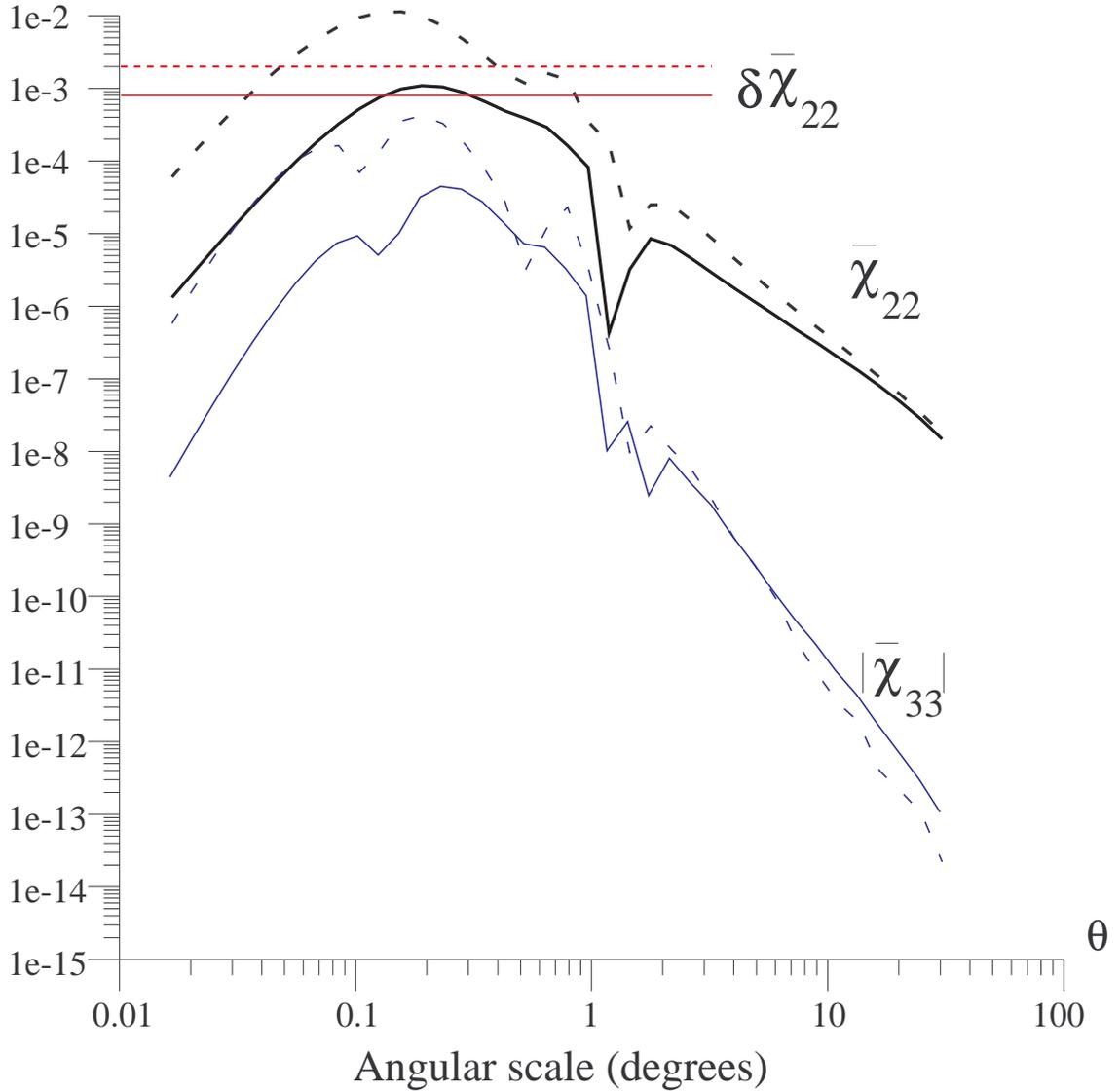}
\label{Fig:Plots}
\caption{The dimensionless cumulants $\bar{\chi}_{22}\left( \theta \right) $
and $\bar{\chi}_{33}\left( \theta \right) $ as functions of the angular
scale $\theta $. The cosmological parameters are: for solid lines
$\Omega_b= 0.05$, $\Omega_\Lambda= 0.3$, $h=0.5$, $n_s=1.2$ (open); for
dashed lines $\Omega_b= 0.03$, $\Omega_\Lambda= 0.97$, $h=0.7$, $n_s=1.0$ (flat).
The horizontal lines represent the standard deviation of $\bar{\chi}_{22}$.}
\end{figure}

\end{document}